\documentclass[12pt]{article}
\usepackage{epsf}
\usepackage{graphicx}
\usepackage{amssymb}
\def\uu{\langle \bar u u \rangle}
\def\dd{\langle \bar d d \rangle}
\def\ss{\langle \bar s s \rangle}

\title{ {${\Sigma_{Q}\Lambda_{Q}\pi}$ Coupling Constant in  Light Cone QCD Sum Rules }}
\author{
   K. Azizi \thanks{ e-mail:e146342@metu.edu.tr},
 M. Bayar\thanks {e-mail:mbayar@metu.edu.tr},
  A. Ozpineci \thanks {e-mail:ozpineci@metu.edu.tr} \\
  \small Physics
Department, Middle East Technical University, 06531, Ankara, Turkey }
 \date{}
\begin{document}
\setlength{\baselineskip}{26pt} \maketitle
\setlength{\baselineskip}{7mm}
\begin{abstract}
The strong coupling constants $g_{{\Sigma_{Q}\Lambda_{Q}\pi}}$ ($Q=b$ and $c$) are studied in the framework of the light cone QCD sum rules using the most general form of the baryonic currents. 
The predicted coupling constants are used to estimate the decay widths for the $\Sigma_{Q}\longrightarrow\Lambda_{Q}\pi$ decays which are  
compared with the predictions of the other approaches and existing experimental data.
\end{abstract}
\thispagestyle{empty}
\newpage
\setcounter{page}{1}
\section{Introduction}
Recent years have witnessed advances in the heavy baryon spectroscopy, with the discoveries of the heavy baryons involving the $b$ and $c$ quarks. Since the spin of the baryon carries
information on the spin of the heavy quark, the study of the heavy baryons might also
lead us to study the spin effects at the loop level in the standard model. 

To study the meson-baryon couplings, a non-perturbative method is needed. Among all non-perturbative approaches, the QCD sum rules approach 
\cite{R2}-\cite{chernyak} has received special attention to study the properties of hadrons. In the case of the light baryons, 
this method has been successfully applied for calculation of the 
meson-baryon coupling constants. The pion-nucleon coupling constant has been studied in traditional three-point QCD sum rules \cite{R3}-\cite{R11}. The kaon-baryon 
coupling constants have also been calculated in the same framework in  \cite{R12}-\cite{R14}. The latter has also been studied in light cone QCD sum rules (LCQSR) 
in \cite{R15}. The coupling constant for K meson-octet baryons and $\pi$ meson-octet baryons have also been calculated in \cite{altug2} in LCQSR.

The QCD sum rules is also applied to the study of the heavy hadron mass spectrum (see e.g. \cite{Zhang:2008pm}). The masses are also studied in QCD string model
\cite{Grach:2008ij} and 
using quark model in \cite{Buisseret:2008tq, Valcarce:2008dr}. In
\cite{Valcarce:2008dr}, sum rules between the masses of the heavy baryons
derived using the quark model has been analyzed and experimental tests of sum rules for heavy baryon masses have been discussed in \cite{Franklin:2008hx}.
In the present work, using the general form of the current for   $\Sigma_{Q}$ and $\Lambda_{Q}$  baryons, we calculate the $g_{{\Sigma_{Q}\Lambda_{Q}\pi}}$ ($Q=b$ and $c$) coupling constants in the framework of the LCQSR approach. Having computed the  coupling constants, we also evaluate the total decay widths for strong $\Sigma_{Q}\longrightarrow\Lambda_{Q}\pi$ decays and compare our results with the predictions of the relativistic three-quark model (RTQM) \cite{ivanov}, light-front quark model (LFQM) \cite{tawfig} and existing experimental data. The paper encompasses three sections: in the next section, we calculate the LCQSR for the coupling constant $g_{{\Sigma_{Q}\Lambda_{Q}\pi}}$. Section III is devoted to the numerical analysis of the coupling constant  $g_{{\Sigma_{Q}\Lambda_{Q}\pi}}$, our prediction for the total decay rates and discussion.

\section{Light cone QCD sum rules for the coupling constant $g_{{\Sigma_{Q}\Lambda_{Q}\pi}}$  }

To calculate the coupling constant $g_{{\Sigma_{Q}\Lambda_{Q}\pi}}$ in LCQSR, one starts with a suitably chosen correlation function.
In this work, the following correlation functions is chosen:
\begin{equation}\label{T}
\Pi=i\int d^{4}xe^{ipx}\langle \pi(q)\mid {\cal T}\{\eta_{\Lambda_{Q}}(x)\bar{\eta}_{\Sigma_{Q}}(0) \}\mid0\rangle,
\end{equation}
where $\eta_{\Sigma_{Q}}$ and $\eta_{\Lambda_{Q}}$ are the interpolating currents of the heavy baryons $\Sigma_{Q}$ and $\Lambda_{Q}$. 
In this correlator, the  hadrons are represented by their interpolating quark currents.
This correlation function can be calculated in two different ways: 
on the one hand, inserting complete sets of hadronic states into the correlation function, it can be expressed in
terms of hadronic parameters such as the masses, residues and the coupling constants. On the other hand, it can be calculated
in terms of quark-gluon parameters in the deep Euclidean region when $p^2 \rightarrow -\infty$ and $(p+q)^2 \rightarrow - \infty$.
The coupling constant is
determined by matching these  two different representations of the
correlation function and applying double Borel transformation with respect to the momentum of both hadrons to suppress the contributions of the higher states and continuum.

The derivation of the physical (or phenomenological) representation of the correlation function 
follows the same lines as in the case of light hadrons (see e.g. \cite{altug2}). For completeness, we repeat the derivation below.
First, one inserts two complete sets of states between the interpolating currents in (\ref{T}) with quantum numbers of the $\Sigma_{Q}$ and $\Lambda_{Q}$ baryons. 
\begin{eqnarray}\label{T2}
\Pi&=&\frac{\langle0\mid \eta_{\Lambda_{Q}}\mid
\Lambda_{Q}(p_{2})\rangle}{p_{2}^{2}-m_{\Lambda_{Q}}^{2}}\langle \Lambda_{Q}(p_{2}) \pi(q)\mid
\Sigma_{Q}(p_{1})\rangle\frac{\langle \Sigma_{Q}(p_{1})\mid
\eta_{\Sigma_{Q}}\mid 0\rangle}{p_{1}^{2}-m_{\Sigma_{Q}}^{2}}+...,\nonumber\\
\end{eqnarray}
where $p_{1}=p+q$,  $p_{2}=p$, and ... stands for the contributions of higher states and continuum. The vacuum to baryon matrix element of the interpolating currents are defined as
\begin{equation}\label{lambdabey}
\langle0\mid \eta_{B}\mid B(p,s)\rangle=\lambda_{B}u_{B}(p,s),
\end{equation}
where $B=\Sigma_{Q}$ or $\Lambda_{Q}$, $u_{B}(p,s)$ is a spinor describing the baryon $B$ and $\lambda_{B}$ is the  residue of the $B$  baryon. 
The last ingredient is the matrix element $\langle \Lambda_{Q}(p_{2}){\pi}(q)\mid
\Sigma_{Q}(p_{1})\rangle$ which can be parameterized in terms of the coupling constant $g_{{\Sigma_{Q}\Lambda_{Q}\pi}}$ as
\begin{eqnarray}\label{matelpar}
\langle \Lambda_{Q}(p_{2}){\pi}(q)\mid
\Sigma_{Q}(p_{1})\rangle &=&g_{\Sigma_{Q}\Lambda_{Q}\pi}\overline{u}(p_{2})i\gamma_{5}u(p_{1}).\nonumber\\
\end{eqnarray}
Using Eqs. (\ref{T2}-\ref{matelpar}) and summing over the spin of the baryons, the following representation of the correlator for the phenomenological side is obtained:
\begin{eqnarray}\label{final phenpart}
\Pi&=&i\frac{g_{\Sigma_{Q}\Lambda_{Q}\pi}\lambda_{\Lambda_{Q}}\lambda_{\Sigma_{Q}}}{(p_{1}^{2}-m_{\Sigma_{Q}}^{2})(p_{2}^{2}-m_{\Lambda_{Q}}^{2})}
\left[-\not\!p \not\!q\gamma_{5}-m_{\Sigma_{Q}}\not\!q\gamma_{5}\right.\nonumber\\&+&\left.(m_{\Lambda_{Q}}-m_{\Sigma_{Q}})\not\!p\gamma_{5}+(m_{\Sigma_{Q}}m_{\Lambda_{Q}}-p^{2})\gamma_{5}\right].
\end{eqnarray}
Note that, the structures $\not\!p\gamma_{5}$ and $\gamma_5$ have very small coefficient due to the fact that $m_{\Sigma_Q} \simeq m_{\Lambda_Q}$, hence they will not yield
reliable sum rules.

To calculate the representation of the correlation function, Eq. (\ref{T}), from QCD side, we need the explicit expressions of the interpolating currents for
$\Sigma_{Q}$ and $\Lambda_{Q}$ baryons. In principal, any operator having the same quantum numbers as the corresponding baryon can be used.
It is well known that there is a continuum of 
choices for the heavy spin-$\frac{1}{2}$ baryons interpolating currents that does not contain any derivatives. The general form of
the $\Sigma_Q$ and $\Lambda_Q$ currents can be written as (see also \cite{chungcurrent})
\begin{eqnarray}\label{currentguy}
\eta_{\Sigma_{Q}}&=&-\frac{1}{\sqrt{2}}\epsilon_{abc}\left\{\vphantom{\int_0^{x_2}}(u_{}^{aT}CQ^{b})\gamma_{5}d^{c
}+\beta(u^{aT}C\gamma_{5}Q^{b})d^{c} \right.\nonumber\\&-& \left. [(Q_{}^{aT}Cd^{b})\gamma_{5}u^{c}+\beta(Q_{}^{aT}C\gamma_{5}d^{b})u^{c}]\right\},\nonumber\\
\eta_{\Lambda_{Q}}&=&\frac{1}{\sqrt{6}}\epsilon_{abc}\left\{\vphantom{\int_0^{x_2}}2[(u_{}^{aT}Cd^{b})\gamma_{5}Q^{c}
+\beta'(u^{aT}C\gamma_{5}d^{b})Q^{c}]+(u_{}^{aT}CQ^{b})\gamma_{5}d^{c}\right.\nonumber\\
&+&\left. \beta'(u_{}^{aT}C\gamma_{5}Q^{b})d^{c}+(Q_{}^{aT}Cd^{b})\gamma_{5}u^{c}+\beta'(Q_{}^{aT}C\gamma_{5}d^{b})u^{c}\right\},\nonumber\\
\end{eqnarray}
where $\beta$ and $\beta'$ are arbitrary parameters. For simplicity, we assume  $\beta=\beta'$. The $\beta=-1$ corresponds to the Ioffe current and $C$ is the charge conjugation operator and  $a$, $b$ and $c$ are color
indices.

After contracting out all quark pairs in  Eq. (\ref{T}), the following expression for the correlation
function in terms of the quark propagators is obtained
\begin{eqnarray}\label{tree expresion.m}
\Pi&=&\frac{i}{\sqrt{3}}\epsilon_{abc}\epsilon_{a'b'c'}\int
d^{4}xe^{ipx}\langle\pi(q)\mid\left\{\vphantom{\int_0^{x_2}}\gamma_{5}S_{Q}^{ca'}
S'^{ab'}_{u}S_{d}^{bc'}\gamma_{5}\right.\nonumber\\&-&\gamma_{5}S_{Q}^{cb'}
S'^{ba'}_{d}S_{u}^{ac'}\gamma_{5}-1/2(\gamma_{5}S_{d}^{ca'}
S'^{bb'}_{Q}S_{u}^{ac'}\gamma_{5}-\gamma_{5}S_{u}^{cb'}
S'^{aa'}_{Q}S_{d}^{bc'}\gamma_{5}\nonumber\\&+&Tr[S_{Q}^{ba'}S'^{ab'}_{u}]\gamma_{5}S^{cc'}_{d}\gamma_{5}-Tr[S_{d}^{ba'}S'^{ab'}_{Q}]\gamma_{5}S^{cc'}_{u}\gamma_{5})\nonumber\\&+& \beta\left[\vphantom{\int_0^{x_2}}\gamma_{5}S_{Q}^{ca'}
\gamma_{5}S'^{ab'}_{u}S_{d}^{bc'}-\gamma_{5}S_{Q}^{cb'}
\gamma_{5}S'^{ba'}_{d}S_{u}^{ac'}+S_{Q}^{ca'}
S'^{ab'}_{u}\gamma_{5}S_{d}^{bc'}\gamma_{5}\right.\nonumber\\&-&S_{Q}^{cb'}
S'^{ba'}_{d}\gamma_{5}S_{u}^{ac'}\gamma_{5}+1/2(\gamma_{5}S_{u}^{cb'}
\gamma_{5}S'^{aa'}_{Q}S_{d}^{bc'}-\gamma_{5}S_{d}^{ca'}
\gamma_{5}S'^{bb'}_{Q}S_{u}^{ac'}\nonumber\\&-&S_{d}^{ca'}
S'^{bb'}_{Q}\gamma_{5}S_{u}^{ac'}\gamma_{5}+S_{u}^{cb'}
S'^{aa'}_{Q}\gamma_{5}S_{d}^{bc'}\gamma_{5}-S^{cc'}_{d}\gamma_{5}Tr[\gamma_{5}S_{Q}^{ba'}S'^{ab'}_{u}]\nonumber\\&+&S^{cc'}_{u}\gamma_{5}Tr[\gamma_{5}S_{d}^{ba'}S'^{ab'}_{Q}]-\gamma_{5}S^{cc'}_{d}Tr[S_{Q}^{ba'}\gamma_{5}S'^{ab'}_{u}]+\left.\gamma_{5}S^{cc'}_{u}Tr[S_{d}^{ba'}\gamma_{5}S'^{ab'}_{Q}])\vphantom{\int_0^{x_2}}\right]\nonumber\\&+&\beta^{2}\left[\vphantom{\int_0^{x_2}}S_{Q}^{ca'}\gamma_{5}
S'^{ab'}_{u}\gamma_{5}S_{d}^{bc'}+S_{Q}^{cb'}\gamma_{5}
S'^{ba'}_{d}\gamma_{5}S_{u}^{ac'}+1/2(S_{u}^{cb'}\gamma_{5}
S'^{aa'}_{Q}\gamma_{5}S_{d}^{bc'}\right.\nonumber\\&-&S_{d}^{ca'}\gamma_{5}
S'^{bb'}_{Q}\gamma_{5}S_{u}^{ac'}-Tr[\gamma_{5}S_{u}^{ab'}\gamma_{5}S'^{ba'}_{Q}]S^{cc'}_{d}+\left.\left.Tr[\gamma_{5}S_{Q}^{ab'}\gamma_{5}S'^{ba'}_{d}]S^{cc'}_{u})\vphantom{\int_0^{x_2}}\right]\vphantom{\int_0^{x_2}}\right\} \mid 0\rangle,\nonumber\\
\end{eqnarray}
where $S'=CS^TC$ and $S_{Q(q)}~~(q=u,d)$ is the full heavy (light) quark
propagator. Note that, the Eq. (\ref{tree expresion.m}) is a schematical representation for the full expression. To obtain the full expression from the Eq. (\ref{tree expresion.m}),
one should replace $S_u$ by $u(0) \bar u(x)$ to calculate the emission from the $u$ quark, and then add to this the result obtained by replacing
$S_d$ by $d(0) \bar d(x)$. From Eq. (\ref{tree expresion.m}), it follows that the expression of the  light and heavy quark propagators are needed.

The light cone expansion of the quark propagator in the external field
is calculated  in \cite{Balitsky}. The propagator  receives contributions
from higher Fock states proportional to the condensates of the operators $\bar q G q$, $\bar q G G q$ and $\bar q q \bar q q$, where $G$ is the gluon field strength tensor. 
In this work, we neglect contributions with two gluons  as well as four quark
operators due to the fact that their contributions are small
 \cite{Braun2}. In this approximation,  the heavy  and
light quark propagators have the  following expressions:
\begin{eqnarray}\label{heavylightguy}
 S_Q (x)& =&  S_Q^{free} (x) - i g_s \int \frac{d^4 k}{(2\pi)^4}
e^{-ikx} \int_0^1 dv \Bigg[\frac{\not\!k + m_Q}{( m_Q^2-k^2)^2}
G^{\mu\nu}(vx)
\sigma_{\mu\nu} \nonumber \\
&+& \frac{1}{m_Q^2-k^2} v x_\mu G^{\mu\nu} \gamma_\nu \Bigg],
\nonumber \\
S_q(x) &=&  S_q^{free} (x)  - \frac{\langle
\bar q q \rangle}{12} -
\frac{x^2}{192} m_0^2 \langle \bar q q \rangle  \nonumber \\ &&
 - i g_s \int_0^1 du \left[\frac{\not\!x}{16 \pi^2 x^2} G_{\mu \nu} (ux) \sigma_{\mu \nu} - u x^\mu
G_{\mu \nu} (ux) \gamma^\nu \frac{i}{4 \pi^2 x^2} \right].
 \end{eqnarray}

 The expression of the free light and heavy quark propagators in the $x$ representation are:

\begin{eqnarray}\label{free1guy}
S^{free}_{q} &=&\frac{i\not\!x}{2\pi^{2}x^{4}},\nonumber\\
S^{free}_{Q}
&=&\frac{m_{Q}^{2}}{4\pi^{2}}\frac{K_{1}(m_{Q}\sqrt{-x^2})}{\sqrt{-x^2}}-i
\frac{m_{Q}^{2}\not\!x}{4\pi^{2}x^2}K_{2}(m_{Q}\sqrt{-x^2}),\nonumber\\
\end{eqnarray}
where $K_{i}$ are the Bessel functions.

 In order to calculate the contributions of the pion emission,  the matrix
 elements $ \langle\pi(q)\mid\bar q
 \Gamma_{i}q\mid0\rangle$ are needed. 
 Here, $\Gamma_{i}$  is any member of the complete set of Dirac matrices 
 $\{1, \gamma_{5}, \gamma_{\alpha}, i\gamma_{5}\gamma_{\alpha}, \sigma_{\alpha\beta}/\sqrt{2}\}$. These matrix  elements are determined  in terms of
 the pion distribution amplitudes (DA's) as follows  \cite{R21,R22}.
\begin{eqnarray}
\langle {\pi}(p)| \bar q(x) \gamma_\mu \gamma_5 q(0)| 0 \rangle &=& -i f_{\pi} p_\mu  \int_0^1 du  e^{i \bar u p x} 
	\left( \varphi_{\pi}(u) + \frac{1}{16} m_{\pi}^2 x^2 {\mathbb A}(u) \right)
\nonumber \\
	&-& \frac{i}{2} f_{\pi} m_{\pi}^2 \frac{x_\mu}{px} \int_0^1 du e^{i \bar u px} {\mathbb B}(u), 
\nonumber \\
\langle {\pi}(p)| \bar q(x) i \gamma_5 q(0)| 0 \rangle &=& \mu_{\pi} \int_0^1 du e^{i \bar u px} \varphi_P(u),
\nonumber \\
\langle {\pi}(p)| \bar q(x) \sigma_{\alpha \beta} \gamma_5 q(0)| 0 \rangle &=& 
\frac{i}{6} \mu_{\pi} \left( 1 - \tilde \mu_{\pi}^2 \right) \left( p_\alpha x_\beta - p_\beta x_\alpha\right)
	\int_0^1 du e^{i \bar u px} \varphi_\sigma(u),
\nonumber \\
\langle {\pi}(p)| \bar q(x) \sigma_{\mu \nu} \gamma_5 g_s G_{\alpha \beta}(v x) q(0)| 0 \rangle &=&
	i \mu_{\pi} \left[
		p_\alpha p_\mu \left( g_{\nu \beta} - \frac{1}{px}(p_\nu x_\beta + p_\beta x_\nu) \right) 
\right. \nonumber \\
	&-&	p_\alpha p_\nu \left( g_{\mu \beta} - \frac{1}{px}(p_\mu x_\beta + p_\beta x_\mu) \right) 
\nonumber \\
	&-&	p_\beta p_\mu \left( g_{\nu \alpha} - \frac{1}{px}(p_\nu x_\alpha + p_\alpha x_\nu) \right)
\nonumber \\ 
	&+&	p_\beta p_\nu \left. \left( g_{\mu \alpha} - \frac{1}{px}(p_\mu x_\alpha + p_\alpha x_\mu) \right)
		\right]
\nonumber \\
	&\times& \int {\cal D} \alpha e^{i (\alpha_{\bar q} + v \alpha_g) px} {\cal T}(\alpha_i),
\nonumber \\
\langle {\pi}(p)| \bar q(x) \gamma_\mu \gamma_5 g_s G_{\alpha \beta} (v x) q(0)| 0 \rangle &=& 
	p_\mu (p_\alpha x_\beta - p_\beta x_\alpha) \frac{1}{px} f_{\pi} m_{\pi}^2 
		\int {\cal D}\alpha e^{i (\alpha_{\bar q} + v \alpha_g) px} {\cal A}_\parallel (\alpha_i)
\nonumber \\
	&+& \left[
		p_\beta \left( g_{\mu \alpha} - \frac{1}{px}(p_\mu x_\alpha + p_\alpha x_\mu) \right) \right.
\nonumber \\
	&-& 	p_\alpha \left. \left(g_{\mu \beta}  - \frac{1}{px}(p_\mu x_\beta + p_\beta x_\mu) \right) \right]
	f_{\pi} m_{\pi}^2
\nonumber \\
	&\times& \int {\cal D}\alpha e^{i (\alpha_{\bar q} + v \alpha _g) p x} {\cal A}_\perp(\alpha_i),
\nonumber \\
\langle {\pi}(p)| \bar q(x) \gamma_\mu i g_s G_{\alpha \beta} (v x) q(0)| 0 \rangle &=& 
	p_\mu (p_\alpha x_\beta - p_\beta x_\alpha) \frac{1}{px} f_{\pi} m_{\pi}^2 
		\int {\cal D}\alpha e^{i (\alpha_{\bar q} + v \alpha_g) px} {\cal V}_\parallel (\alpha_i)
\nonumber \\
	&+& \left[
		p_\beta \left( g_{\mu \alpha} - \frac{1}{px}(p_\mu x_\alpha + p_\alpha x_\mu) \right) \right.
\nonumber \\
	&-& 	p_\alpha \left. \left(g_{\mu \beta}  - \frac{1}{px}(p_\mu x_\beta + p_\beta x_\mu) \right) \right]
	f_{\pi} m_{\pi}^2
\nonumber \\
	&\times& \int {\cal D}\alpha e^{i (\alpha_{\bar q} + v \alpha _g) p x} {\cal V}_\perp(\alpha_i),
\end{eqnarray}
where $\mu_{\pi} = f_{\pi} \frac{m_{\pi}^2}{m_{u} + m_{d}},$ 
$\tilde \mu_{\pi} = \frac{{m_{u} + m_{d}}}{m_{\pi}}$ and
the functions $\varphi_{\pi}(u),$ $\mathbb A(u),$ $\mathbb B(u),$ $\varphi_P(u),$ $\varphi_\sigma(u),$ 
${\cal T}(\alpha_i),$ ${\cal A}_\perp(\alpha_i),$ ${\cal A}_\parallel(\alpha_i),$ ${\cal V}_\perp(\alpha_i)$ and ${\cal V}_\parallel(\alpha_i)$
are functions of definite twist and their expressions will be given in the numerical analysis section. The measure $\cal D\alpha$ is defined as
\begin{equation}
\int {\cal D} \alpha = \int_0^1 d \alpha_{\bar q} \int_0^1 d
\alpha_q \int_0^1 d \alpha_g \delta(1-\alpha_{\bar
q}-\alpha_q-\alpha_g).\nonumber \\
\end{equation}
Note that, in the approximation of this work where we neglect the light
quark masses, $m_\pi^2=0$, $\tilde\mu_\pi = 0$, $\mu_\pi = - \uu / f_\pi =
- \dd / f_\pi$.

Using the expressions of the light and heavy full  propagators and the pion DA's, the correlation function Eq. (\ref{T}) can be calculated in terms of QCD parameters.
Separating the coefficient of the structure
$ \not\!p\not\!q\gamma_{5} $ in both representations,
and equating them,  sum rules for
the coupling constant $g_{{\Sigma_{Q}\Lambda_{Q}\pi}}$ is obtained. The contribution of the higher states is subtracted using quark hadron duality, and in order to 
further suppress their contribution, Borel transformation with respect to the variables $p_{2}^2=p^2$ and $p_{1}^2=(p+q)^2$
 is applied. Here, we should mention that we have also studied the other structure in Eq. (\ref{final phenpart}), i.e., $\not\!q\gamma_{5}$ but its result for coupling constant is not stable and only the $ \not\!p\not\!q\gamma_{5} $ structure leads to reliable prediction on the coupling constant $g_{{\Sigma_{Q}\Lambda_{Q}\pi}}$.

The sum rules for the coupling constant is obtained as
\begin{eqnarray}\label{magneticmoment1}
\lambda_{\Sigma_{Q}}\lambda_{\Lambda_{Q}}e^{-\frac{m_{\Lambda_{Q}}^{2}+m_{\Sigma_{Q}}^{2}}{2M^{2}}}g_{\Sigma_{Q}\Lambda_{Q}\pi}=\Pi,
\end{eqnarray}
where the function $\Pi$ is
\begin{eqnarray}\label{magneticmoment2}
\Pi&=&\int_{m_{Q}^{2}}^{s_{0}}e^{\frac{-s}{M^{2}}}\rho(s)ds+e^{\frac{-m_Q^2}{M^{2}}}\Gamma,
\end{eqnarray}
with
\begin{eqnarray}\label{rho1}
\rho(s)&=&(<\overline{d}d>+<\overline{u}u>)\frac{1}{12\sqrt{6}}f_{\pi}(\beta-1)\beta \varphi_{\pi}(u_{0})\psi_{00}
\nonumber\\&-&\frac{1}{96\sqrt{6}\pi^{2}}\Bigg[m_{Q}(\beta-1)\Bigg\{-6\Bigg[-2(\psi_{20}-\psi_{31})m_{Q} \mu_{\pi}[-\zeta_{5}(1+2\beta)+\zeta_{6}(1+\beta)]\nonumber\\&-&
\psi_{10}m_{Q} \mu_{\pi}[3\zeta_{5}(1+\beta)-4\zeta_{6}]-m_{Q}\mu_{\pi}(3\zeta_{5}(1+\beta)-4\zeta_{6})ln(\frac{m_Q^{2}}{s})\Bigg]\nonumber\\&+& 
6f_{\pi}m_{Q}^{2}(1+\beta)[2\psi_{10}-\psi_{20}
+\psi_{31}+2ln(\frac{m_Q^{2}}{s})\varphi_{\pi}(u_{0})\nonumber\\&+& 2(\psi_{20}-\psi_{31})m_{Q}\mu_{\pi}(1+2\beta)\varphi_{\sigma}(u_{0})
\Bigg\}\Bigg],
\end{eqnarray}
and
\begin{eqnarray}\label{gamma1}
\Gamma&=&\frac{m_{0}^{2}}{192\sqrt{6}}(<\overline{d}d>+<\overline{u}u>)\left[\vphantom{\int_0^{x_2}}
\right.
\frac{2 f_{\pi}}{9}[-11-17\beta+(7+\beta)](\beta-1)\varphi_{\pi}(u_{0})\nonumber\\&-&
\frac{8m_{Q}^{2}}{3M^{2}}m_{Q}\mu_{\pi}(\beta^{2}+\beta+1)\varphi_{\sigma}(u_{0})
\nonumber\\&-&\frac{4 m_Q}{9 M^{2}}\Bigg\{9f_{\pi}m_{Q}(\beta-1)\beta\varphi_{\pi}(u_{0})-\mu_{\pi}(3\beta^{2}+2\beta+3)\varphi_{\sigma}(u_{0})\Bigg\}\Bigg]\nonumber\\&+&\frac{1}{6\sqrt{6}}(<\overline{d}d>+<\overline{u}u>)m_{Q}\mu_{\pi}(\beta^{2}+\beta+1)\varphi_{\sigma}(u_{0}).
\end{eqnarray}
The other functions entering Eqs. (\ref{rho1}-\ref{gamma1}) are given as
\begin{eqnarray}\label{etalar}
\zeta_{j} &=& \int {\cal D}\alpha_i \int_0^1 dv f_{j}(\alpha_i)
\delta(\alpha_{ q} + v \alpha_g -  u_0),
\nonumber \\
\zeta'_{j} &=& \int {\cal D}\alpha_i \int_0^1 dv g_{j}(\alpha_i)
\delta'(\alpha_{ q} + v \alpha_g -  u_0),
\nonumber \\
\psi_{nm}&=&\frac{{( {s-m_{Q}}^2 )
}^n}{s^m{(m_{Q}^{2})}^{n-m}},\nonumber \\
\end{eqnarray}
 and  $f_{1}(\alpha_i)={\cal V_{\parallel}}(\alpha_i)$, $f_{2}(\alpha_i)=v{\cal V_{\parallel}}(\alpha_i)$, $f_{3}(\alpha_i)={\cal V_{\perp}}(\alpha_i)$, $f_{4}(\alpha_i)=v{\cal V_{\perp}}(\alpha_i)$, $g_{1}(\alpha_i)={\cal T}(\alpha_i)$ and  $g_{2}(\alpha_i)=v{\cal T}(\alpha_i)$ are the pion distribution amplitudes. Note that, in the above equations, the Borel parameter $M^2$  is defined as $M^{2}=\frac{M_{1}^{2}M_{2}^{2}}{M_{1}^{2}+M_{2}^{2}}$ and
$u_{0}=\frac{M_{1}^{2}}{M_{1}^{2}+M_{2}^{2}}$.  Since the mass of the initial and final baryons are close to each other,
 we can set $ M_{1}^{2} = M_{2}^{2} = 2M^{ 2}$ and $u_{0} =\frac{1}{2}$. The contributions of the terms $\sim <G^2>$ are also calculated, but their numerical values are very small and therefore for customary in the expressions these terms are omitted.

For calculation of the coupling constants of the considered baryons,
their residues, $\lambda_{\Sigma_{Q}(\Lambda_{Q})}$ are needed. Their expressions are obtained as:
\begin{eqnarray}\label{residu2}
-\lambda_{\Sigma_{Q}(\Lambda_{Q})}^{2}e^{-m_{\Sigma_{Q}(\Lambda_{Q})}^{2}/M^{2}}&=&\int_{m_{Q}^{2}}^{s_{0}}e^{\frac{-s}{M^{2}}}\rho_{1(2)}(s)ds+e^{\frac{-m_Q^2}{M^{2}}}\Gamma_{1(2)},
\end{eqnarray}
with
\begin{eqnarray}\label{residurho1}
\rho_{1}(s)&=&(<\overline{d}d>+<\overline{u}u>)\frac{(\beta^{2}-1)}{64 \pi^{2}}\Bigg\{\frac{m_{0}^{2}}{4 m_{Q}}(6\psi_{00}-13\psi_{02}-6\psi_{11})\nonumber\\&+&3m_{Q}(2\psi_{10}-\psi_{11}-\psi_{12}+2\psi_{21})\Bigg\}\nonumber\\&+&\frac{ m_{Q}^{4}}{2048 \pi^{4}} [5+\beta(2+5\beta)][12\psi_{10}-6\psi_{20}+2\psi_{30}-4\psi_{41}+\psi_{42}-12 ln(\frac{s}{m_{Q}^{2}})],\nonumber\\
\end{eqnarray}
\begin{eqnarray}\label{residurho2}
\rho_{2}(s)&=&(<\overline{d}d>+<\overline{u}u>)\frac{(\beta-1)}{192 \pi^{2}}\Bigg\{\frac{m_{0}^{2}}{4 m_{Q}}[6(1+\beta)\psi_{00}-(7+11\beta)\psi_{02}\nonumber\\&-&6(1+\beta)\psi_{11}]+(1+5\beta)m_{Q}(2\psi_{10}-\psi_{11}-\psi_{12}+2\psi_{21})\Bigg\}\nonumber\\&+&\frac{ m_{Q}^{4}}{2048 \pi^{4}} [5+\beta(2+5\beta)][12\psi_{10}-6\psi_{20}+2\psi_{30}-4\psi_{41}+\psi_{42}-12 ln(\frac{s}{m_{Q}^{2}})],\nonumber\\
\end{eqnarray}
\begin{eqnarray}\label{lamgamma1}
\Gamma_{1}&=&\frac{ (\beta-1)^{2}}{24}<\overline{d}d><\overline{u}u>\left[\vphantom{\int_0^{x_2}}\right.\frac{m_{Q}^{2}m_{0}^{2}}{2 M^{4}}+\frac{m_{0}^{2}}{4 M^{2}}-1\Bigg],\nonumber\\
\Gamma_{2}&=&\frac{ (\beta-1)}{72}<\overline{d}d><\overline{u}u>\left[\vphantom{\int_0^{x_2}}\right.\frac{m_{Q}^{2}m_{0}^{2}}{2 M^{4}}(13+11\beta)\nonumber\\&+&\frac{m_{0}^{2}}{4 M^{2}}(25+23\beta)-(13+11\beta)\Bigg].
\end{eqnarray}

%

\section{Numerical analysis}
This section is devoted to the numerical analysis for the coupling constant $g_{{\Sigma_{Q}\Lambda_{Q}\pi}}$ and calculation of the total decay width  for $\Sigma_{Q}\longrightarrow\Lambda_{Q}\pi$. The input parameters used in the analysis of the sum rules are $\uu(1~GeV) = \dd(1~GeV)= -(0.243)^3~GeV^3$, $\ss(1~GeV) = 0.8
\uu(1~GeV)$, $m_b = 4.7~GeV$,  $m_c = 1.23~GeV$, $m_{\Sigma_{b}} = 5.805~GeV$, $m_{\Sigma_{c}} = 2.439~GeV$, $m_{\Lambda_{b}} = 5.622~GeV$, $m_{\Lambda_{c}} = 2.297~GeV$, and $m_0^2(1~GeV) = (0.8\pm0.2)~GeV^2$ \cite{Belyaev}.
From the sum rules for coupling constant, it is clear that the $\pi$-meson wave functions are needed. These wave functions are given as \cite{R21, R22}
\begin{eqnarray}
\phi_{\pi}(u) &=& 6 u \bar u \left( 1 + a_1^{\pi} C_1(2 u -1) + a_2^{\pi} C_2^{3 \over 2}(2 u - 1) \right), 
\nonumber \\
{\cal T}(\alpha_i) &=& 360 \eta_3 \alpha_{\bar q} \alpha_q \alpha_g^2 \left( 1 + w_3 \frac12 (7 \alpha_g-3) \right),
\nonumber \\
\phi_P(u) &=& 1 + \left( 30 \eta_3 - \frac{5}{2} \frac{1}{\mu_{\pi}^2}\right) C_2^{1 \over 2}(2 u - 1) 
\nonumber \\ 
&+&	\left( -3 \eta_3 w_3  - \frac{27}{20} \frac{1}{\mu_{\pi}^2} - \frac{81}{10} \frac{1}{\mu_{\pi}^2} a_2^{\pi} \right) C_4^{1\over2}(2u-1),
\nonumber \\
\phi_\sigma(u) &=& 6 u \bar u \left[ 1 + \left(5 \eta_3 - \frac12 \eta_3 w_3 - \frac{7}{20}  \mu_{\pi}^2 - \frac{3}{5} \mu_{\pi}^2 a_2^{\pi} \right)
C_2^{3\over2}(2u-1) \right],
\nonumber \\
{\cal V}_\parallel(\alpha_i) &=& 120 \alpha_q \alpha_{\bar q} \alpha_g \left( v_{00} + v_{10} (3 \alpha_g -1) \right),
\nonumber \\
{\cal A}_\parallel(\alpha_i) &=& 120 \alpha_q \alpha_{\bar q} \alpha_g \left( 0 + a_{10} (\alpha_q - \alpha_{\bar q}) \right),
\nonumber \\
{\cal V}_\perp (\alpha_i) &=& - 30 \alpha_g^2\left[ h_{00}(1-\alpha_g) + h_{01} (\alpha_g(1-\alpha_g)- 6 \alpha_q \alpha_{\bar q}) +
	h_{10}(\alpha_g(1-\alpha_g) - \frac32 (\alpha_{\bar q}^2+ \alpha_q^2)) \right],
\nonumber \\
{\cal A}_\perp (\alpha_i) &=& 30 \alpha_g^2(\alpha_{\bar q} - \alpha_q) \left[ h_{00} + h_{01} \alpha_g + \frac12 h_{10}(5 \alpha_g-3) \right],
\nonumber \\
B(u)&=& g_{\pi}(u) - \phi_{\pi}(u),
\nonumber \\
g_{\pi}(u) &=& g_0 C_0^{\frac12}(2 u - 1) + g_2 C_2^{\frac12}(2 u - 1) + g_4 C_4^{\frac12}(2 u - 1),
\nonumber \\
{\mathbb A}(u) &=& 6 u \bar u \left[\frac{16}{15} + \frac{24}{35} a_2^{\pi}+ 20 \eta_3 + \frac{20}{9} \eta_4 +
	\left( - \frac{1}{15}+ \frac{1}{16}- \frac{7}{27}\eta_3 w_3 - \frac{10}{27} \eta_4 \right) C_2^{3 \over 2}(2 u - 1) 
	\right. \nonumber \\ 
	&+& \left. \left( - \frac{11}{210}a_2^{\pi} - \frac{4}{135} \eta_3w_3 \right)C_4^{3 \over 2}(2 u - 1)\right]
\nonumber \\
&+& \left( -\frac{18}{5} a_2^{\pi} + 21 \eta_4 w_4 \right)\left[ 2 u^3 (10 - 15 u + 6 u^2) \ln u 
\right. \nonumber \\
&+& \left. 2 \bar u^3 (10 - 15 \bar u + 6 \bar u ^2) \ln\bar u + u \bar u (2 + 13 u \bar u) \right] 
\label{wavefns},
\end{eqnarray}
where $C_n^k(x)$ are the Gegenbauer polynomials,  
\begin{eqnarray}
h_{00}&=& v_{00} = - \frac13\eta_4,
\nonumber \\
a_{10} &=& \frac{21}{8} \eta_4 w_4 - \frac{9}{20} a_2^{\pi},
\nonumber \\
v_{10} &=& \frac{21}{8} \eta_4 w_4,
\nonumber \\
h_{01} &=& \frac74  \eta_4 w_4  - \frac{3}{20} a_2^{\pi},
\nonumber \\
h_{10} &=& \frac74 \eta_4 w_4 + \frac{3}{20} a_2^{\pi},
\nonumber \\
g_0 &=& 1,
\nonumber \\
g_2 &=& 1 + \frac{18}{7} a_2^{\pi} + 60 \eta_3  + \frac{20}{3} \eta_4,
\nonumber \\
g_4 &=&  - \frac{9}{28} a_2^{\pi} - 6 \eta_3 w_3
\label{param0}.
\end{eqnarray}

The constants appearing in the wave functions are calculated at the renormalization scale $\mu=1 ~~GeV^{2}$ and they are given as
$a_{1}^{\pi} = 0$, $a_{2}^{\pi} = 0.44$,
$\eta_{3} =0.015$, $\eta_{4}=10$, $w_{3} = -3$ and 
$ w_{4}= 0.2$.

The sum rules for the coupling constant also contains three auxiliary
parameters: Borel mass parameter $M^2$, continuum threshold $s_{0}$ and general parameter $\beta$ enters the expressions of the interpolating currents. 
In principal, $M^2$ and $\beta$ are completely arbitrary and hence the coupling constant, which is a physical observable, should be independent of their exact values.
In practice, though, due to the approximations made in the calculations, there is a residual dependence of the predictions on these unphysical parameters. Hence, a range for these
parameter should be found where the predictions are practically insensitive to variations of these parameters.
To find the working region for  $M^2$, we proceed as
follows. The upper bound is obtained requiring that the contribution
of the higher states and continuum should be less than that of the ground state. The lower bound of $M^2$ is determined from condition
that the highest power of $1/M^{2}$ be less than  say $30^0/_{0}$ of
the highest power of $M^{2}$. These two conditions are both
satisfied in the region $15~GeV^2\leq M^{2}\leq30~GeV^2 $ and
$4~GeV^2\leq M^{2}\leq10~GeV^2 $ for baryons containing b and
c-quark, respectively. The third parameter, $s_0$ has a physical meaning, and it should have a value near the first excited state. 
The value of the  continuum threshold is calculated from the two-point sum rules. We choose the interval  $s_{0}=(6.0^{2}-6.2^{2})~~GeV^{2}$ and $s_{0}=(2.5^{2}-2.7^{2})~~GeV^{2}$ for baryons 
containing the $b$ and $c$ quark, respectively.  

In Figs. 1 and 2, we present the dependence of the coupling constants $g_{{\Sigma_{b}\Lambda_{b}\pi}}$ and $g_{{\Sigma_{c}\Lambda_{c}\pi}}$, at fixed values of
the continuum threshold $s_{0}$ and the general parameter $\beta$. From these figures, we see a good stability for coupling constants $g_{{\Sigma_{b}\Lambda_{b}\pi}}$ and $g_{{\Sigma_{c}\Lambda_{c}\pi}}$ with respect to the Borel mass square $M^{2}$ in the working region. The next step is to determine the working region for auxiliary parameter $\beta$. For this aim, in Figs. 3 and 4, we depict the dependence of the coupling constants $g_{{\Sigma_{b}\Lambda_{b}\pi}}$ and $g_{{\Sigma_{c}\Lambda_{c}\pi}}$ on $cos\theta$ where $tan\theta=\beta$, at two fixed values of $M^{2}$. From these Figures, we see that the best stability for the coupling constants $g_{{\Sigma_{b}\Lambda_{b}\pi}}$ and $g_{{\Sigma_{c}\Lambda_{c}\pi}}$ is in the region $-0.5\leq cos\theta \leq0.2 $.

Our final results on coupling constants $g_{{\Sigma_{b}\Lambda_{b}\pi}}$ and $g_{{\Sigma_{c}\Lambda_{c}\pi}}$ are: 
\begin{eqnarray}\label{saydeg}
g_{{\Sigma_{b}\Lambda_{b}\pi}}&=&23.5\pm4.9,\nonumber\\
g_{{\Sigma_{c}\Lambda_{c}\pi}}&=&10.8\pm2.2.\nonumber\\
\end{eqnarray}
The quoted errors are due to the uncertainties in the input parameters as well as variation of the Borel parameter $M^{2}$, continuum threshold $s_{0} $ and general parameter $\beta$.

Having computed the coupling constant $g_{{\Sigma_{Q}\Lambda_{Q}\pi}}$, the next step is to calculate the total decay width for $\Sigma_{b}\longrightarrow\Lambda_{b}\pi$ and $\Sigma_{c}\longrightarrow\Lambda_{c}\pi$ decays.
From Eq. (\ref{matelpar}) the transition amplitude is $M=g_{\Sigma_{Q}\Lambda_{Q}\pi}\overline{u}i\gamma_{5}u$ and the differential decay width is found in terms of the coupling constant as:
\begin{eqnarray}\label{decaywidth}
\Gamma&=&\frac{\vert g_{{\Sigma_{Q}\Lambda_{Q}\pi}}\vert^{2}}{8\pi m_{\Sigma_{Q}}^{2}}(m_{\Sigma_{Q}}-m_{\Lambda_{Q}})^{2}\vert\overrightarrow{q}\vert,\nonumber\\
\end{eqnarray}
where $\vert\overrightarrow{q}\vert=(m_{\Sigma_{Q}}^{2}-m_{\Lambda_{Q}}^{2})/2 m_{\Sigma_{Q}}$. The numerical values of the decay rates are given in Table 1. In order to compare with the predictions of other methods, in the same table, we present the predictions of the   relativistic three-quark model (RTQM) \cite{ivanov}, light-front quark model (LFQM) \cite{tawfig} and existing experimental data  \cite{pdg11}. This table depicts a good consistency among the methods and the experimental data in order of magnitudes for charm case. Note that, due to the isospin symmetry the decays of different charge $\Sigma_{c}^{++,+}\longrightarrow\Lambda_{c}^{+}\pi^{+,0,-}(\Sigma_{b}^{+,0,-}\longrightarrow\Lambda_{b}\pi^{+,0,-})$ have the same decay widths. Experimentally, only the widths for $\Sigma_{c}^{++,0}\longrightarrow\Lambda_{c}^{+}\pi^{+,-}$ are measured and the value in the table is their average. Only the upper bound for the $\Sigma_{c}^{+}\longrightarrow\Lambda_{c}^{+}\pi^{0}$ is known and it is consistent with other decay modes. Our prediction for the decay rate of the bottom case can be tested in the future experiments.

\begin{table}[h]
\centering
\begin{tabular}{|c|c|c|} \hline 
 &$\Gamma(\Sigma_{c}\longrightarrow\Lambda_{c}\pi$) & $\Gamma(\Sigma_{b}\longrightarrow\Lambda_{b}\pi)$  \\\cline{1-3}
Present work  & $2.16\pm0.85$  & $3.93\pm1.5$ \\\cline{1-3}
RTQM \cite{ivanov} &$3.63\pm0.27$ & -  \\\cline{1-3}
LFQM \cite{tawfig} & $1.555\pm0.165$ & -   \\\cline{1-3}
Exp. \cite{pdg11}  &$2.21\pm0.40$&- \\\cline{1-3}
 \end{tabular}
 \vspace{0.8cm}
\caption{Results for the   decay rates of $\Sigma_{Q}\longrightarrow\Lambda_{Q}\pi$ in different approaches in MeV.
}\label{tab:1}
\end{table}

In summary,  we calculated the $g_{{\Sigma_{b}\Lambda_{b}\pi}}$ and $g_{{\Sigma_{c}\Lambda_{c}\pi}}$ coupling constants in the light cone QCD sum rules approach. Using these coupling constants, we also evaluated the total decay width for the strong $\Sigma_{b}\longrightarrow\Lambda_{b}\pi$ and $\Sigma_{c}\longrightarrow\Lambda_{c}\pi$ decays and compared with the predictions of the other approaches and existing experimental data.
\section{Acknowledgment}
The authors thank T. M. Aliev for his useful discussions and  also  TUBITAK,
Turkish Scientific and Research Council, for their partial financial
support through the project number 106T333, and K. A. and A. O. would like to thank TUBA for their partial financial support through GEBIP.

\clearpage
 \begin{figure}[h!]
\begin{center}
\includegraphics[width=9cm]{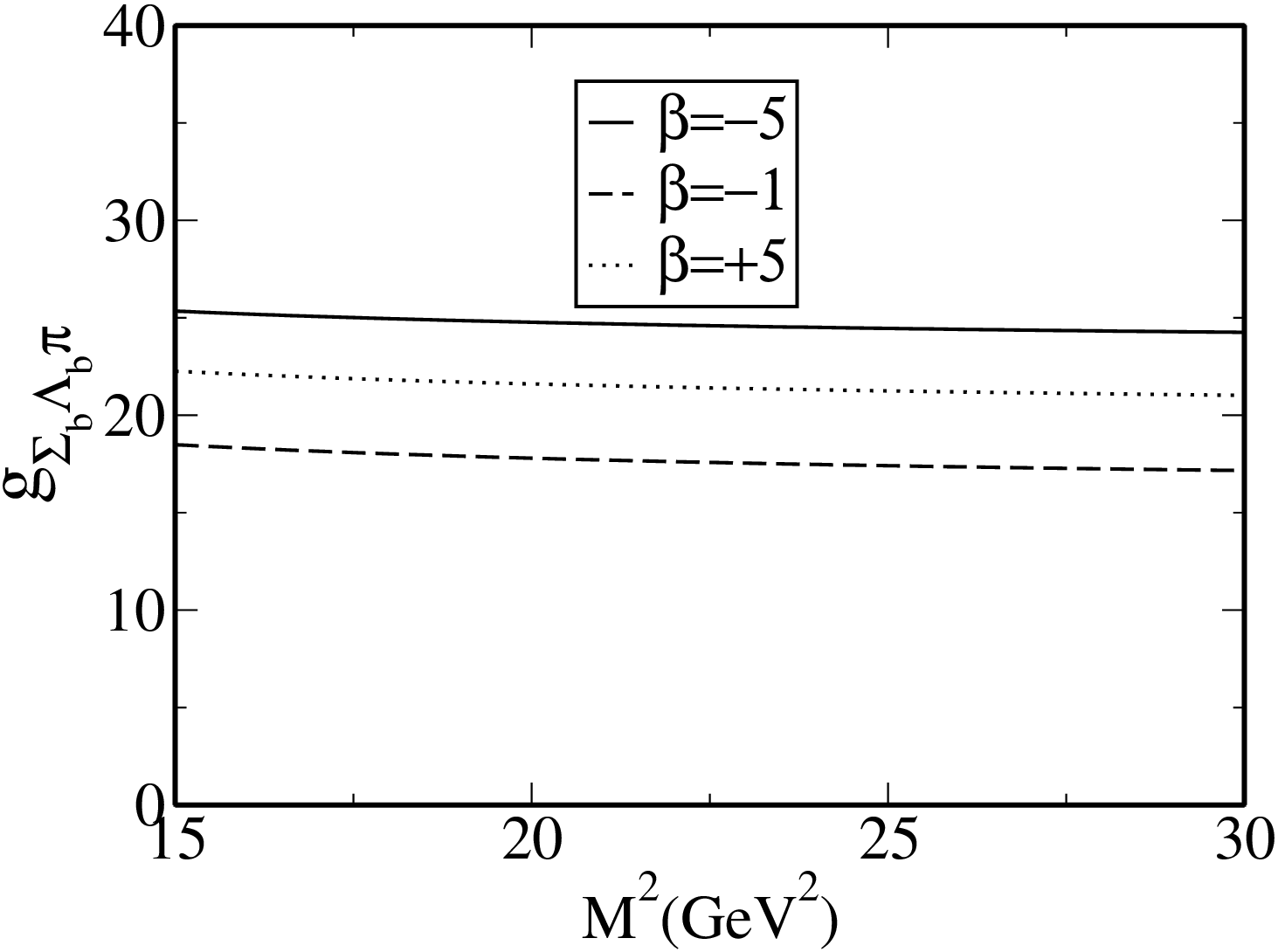}
\end{center}
\caption{The dependence of the $g_{{\Sigma_{b}\Lambda_{b}\pi}}$  on the
Borel parameter $M^{2}$ at fixed value of the continuum
threshold $s_{0}=6.0^2$.} \label{fig1}
\end{figure}
\begin{figure}[h!]
\begin{center}
\includegraphics[width=9cm]{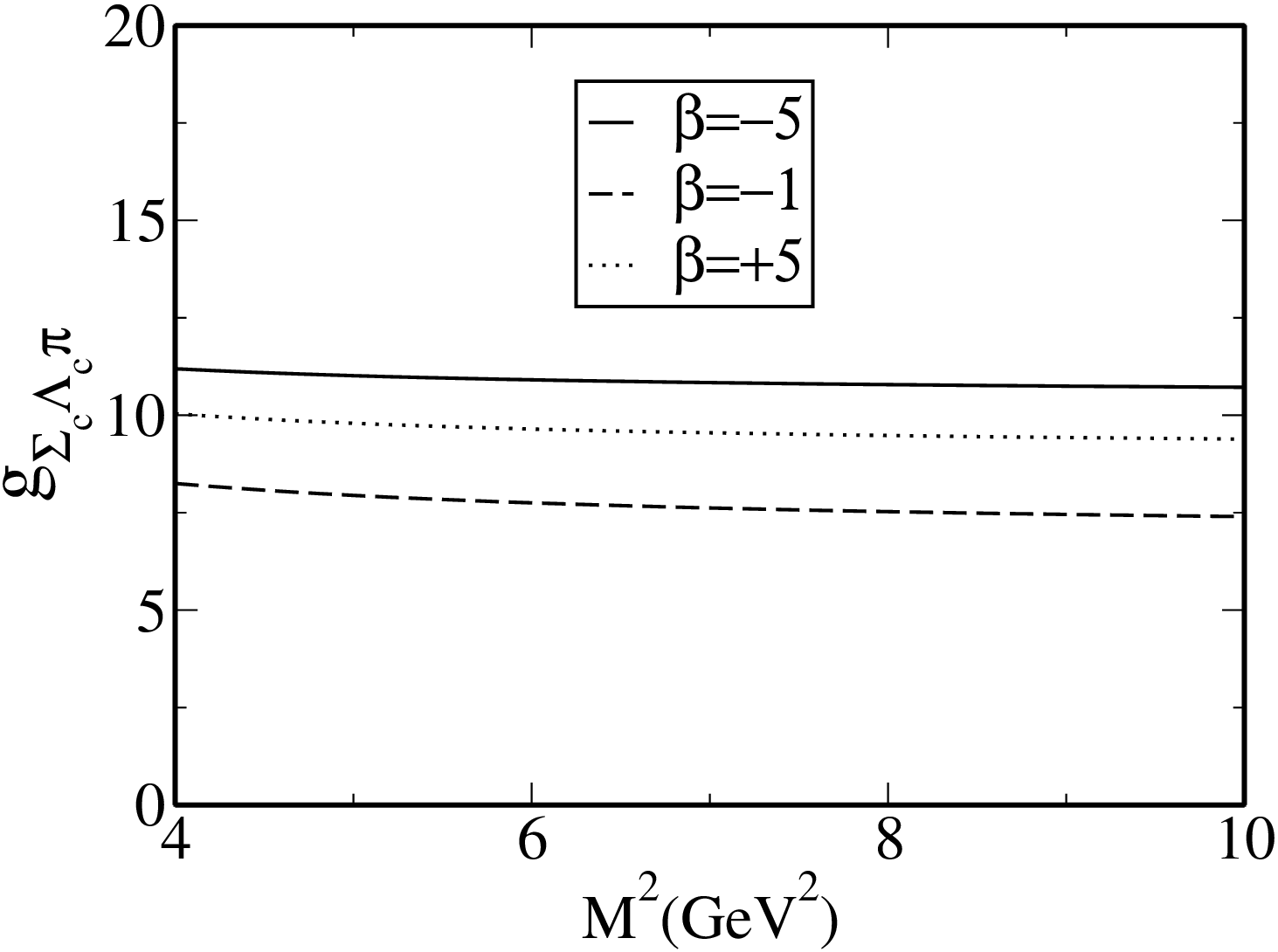}
\end{center}
\caption{The same as Fig. 1, but for $g_{{\Sigma_{c}\Lambda_{c}\pi}}$ and fixed value of the continuum
threshold $s_{0}=2.5^2$.} \label{fig3}
\end{figure}
\begin{figure}[h!]
\begin{center}
\includegraphics[width=9cm]{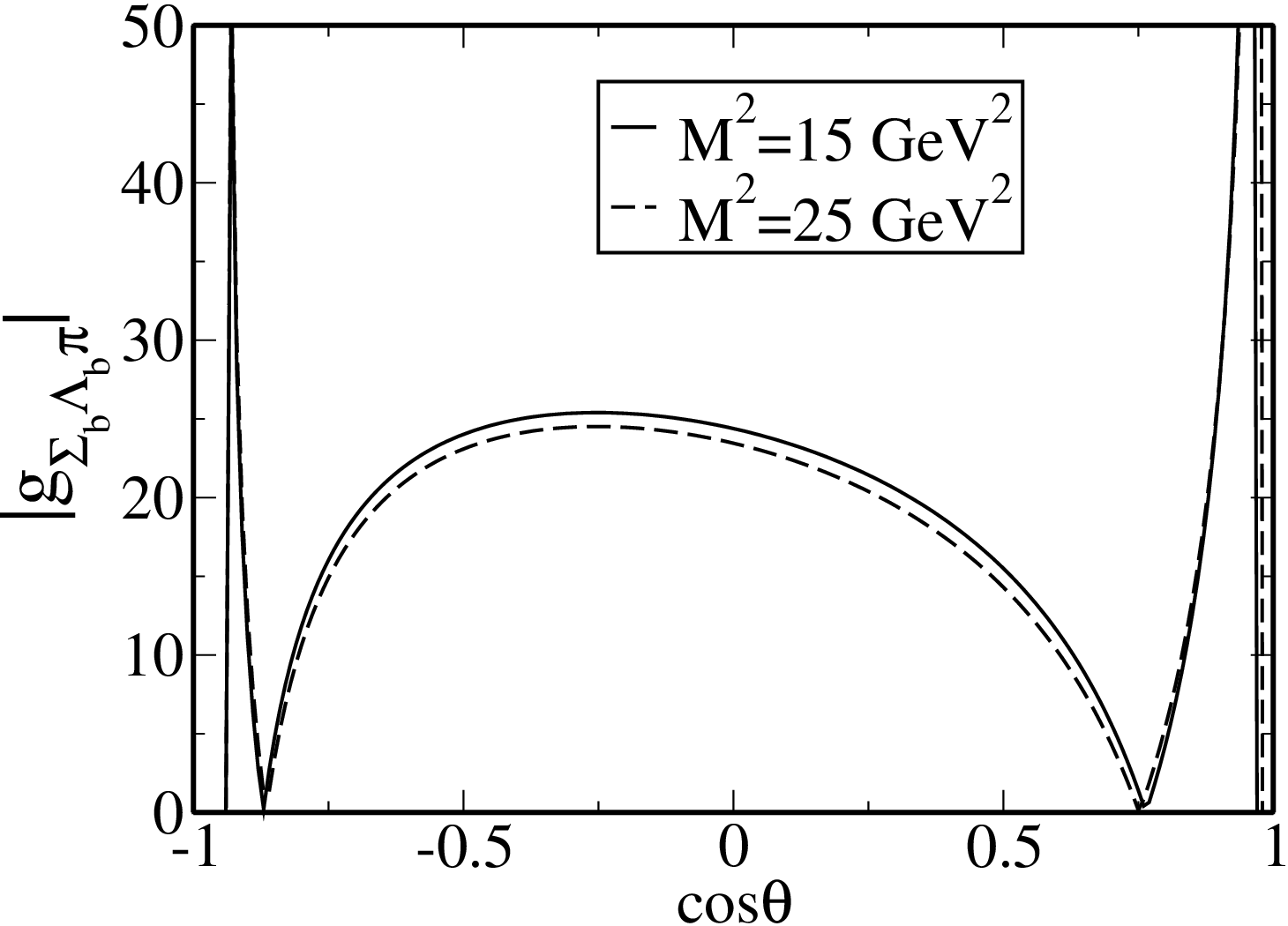}
\end{center}
\caption{The dependence of  $\vert g_{{\Sigma_{b}\Lambda_{b}\pi}}\vert$  on
$cos\theta$ at fixed value of the continuum
threshold $s_{0}=6.0^2$.} \label{fig2}
\end{figure}

\begin{figure}[h!]
\begin{center}
\includegraphics[width=9cm]{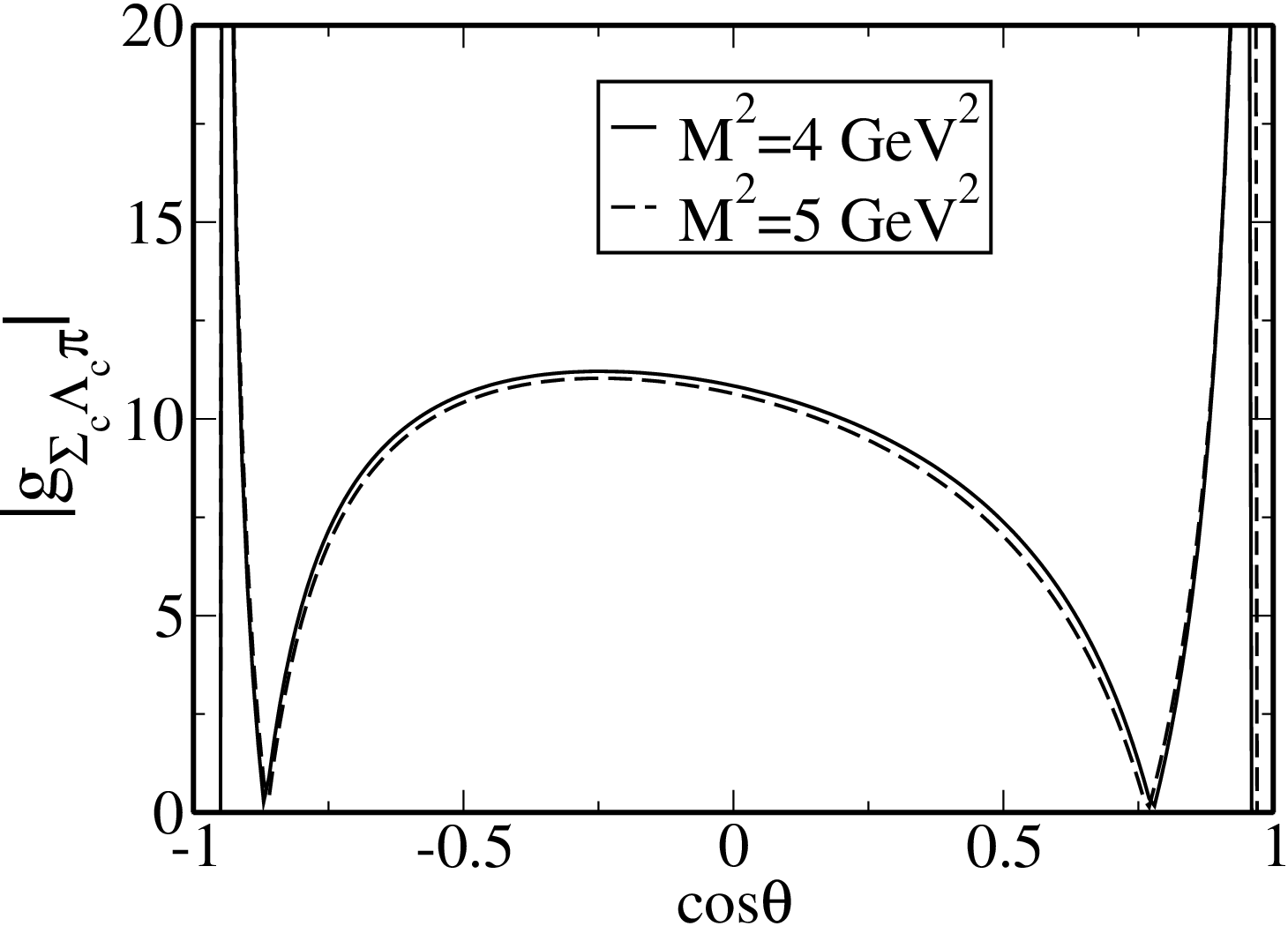}
\end{center}
\caption{The same as Fig. 3, but for  $\vert g_{{\Sigma_{c}\Lambda_{c}\pi}} \vert$ and fixed value of the continuum
threshold $s_{0}=2.5^2$.} \label{fig4}
\end{figure}
\end{document}